\documentclass[11pt,a4paper]{article}
\pdfoutput=1
\usepackage{jcappub}
\usepackage{xcolor}

\newcommand{\pd}{\partial}

\newcommand{\be}{\begin{equation}}
\newcommand{\ee}{\end{equation}}

\title{The sound of DHOST}

\author[a,b]{E.~Babichev,}
\author[a,c]{A.~Leh\'ebel}

\affiliation[a]{Laboratoire de Physique Th\'eorique,
CNRS, Univ. Paris-Sud,\\
Universit\'e Paris-Saclay, F-91405 Orsay, France}
\affiliation[b]{Sorbonne Universit\'e, CNRS, UMR7095,
Institut d'Astrophysique de Paris,
${\mathcal{G}}{\mathbb{R}}\varepsilon{\mathbb{C}}{\mathcal{O}}$,\\
98bis boulevard Arago, F-75014 Paris, France}
\affiliation[c]{School of Physics and Astronomy, University of Nottingham,\\
Nottingham NG7 2RD, United Kingdom
}
\emailAdd{eugeny.babichev@th.u-psud.fr}
\emailAdd{antoine.lehebel@th.u-psud.fr}

\abstract{We show that, in generic higher-order scalar-tensor theories which avoid the Ostrogradsky instability, the presence of a scalar field significantly modifies the propagation of matter perturbations, even in weakly curved backgrounds. This affects notably the speed of sound in the atmosphere of the Earth. It can also generate instabilities in homogeneous media. We use this to constrain the viable higher-order scalar-tensor models.}

\begin{document}

\maketitle

\date{\today}

\section{Introduction}

The most general class of scalar-tensor theories with only one scalar degree of freedom ---~therefore avoiding the Ostrogradsky ghost \cite{Ostrogradski}~--- has been presented recently \cite{Crisostomi:2016czh,Achour:2016rkg} as a maximal extension of Horndeski theory \cite{Horndeski:1974wa}. One of the main motivations for studying such theories, dubbed Degenerate Higher Order Scalar-Tensor (DHOST) theories or Extended Scalar-Tensor (EST) theories, is to modify General Relativity (GR) in order to explain the present-day acceleration of the Universe. However, on a non-trivial cosmological background, gravitational waves generically propagate at a speed which differs from the speed of light, due to the kinetic mixing between the metric and the scalar.
This property, together with recent observations, invalidates generic DHOST models as dark energy candidates. 
Indeed, the observation of gravitational waves generated by a binary neutron star merger, along with their electromagnetic counterpart,  puts tight constraints on the difference between the speeds of gravity and light \cite{TheLIGOScientific:2017qsa,Monitor:2017mdv}. The most general DHOST theories do not satisfy this constraint. However, there is a subfamily of theories where gravity waves propagate with the same speed as light on cosmological background\footnote{It has been shown that, for high curvatures, near black holes in a de Sitter universe, the equality $c_\mathrm{gravity}=c_\mathrm{light}$ still holds \cite{Babichev:2017lmw,Babichev:2018uiw} (when it holds in the asymptotic de Sitter region).}. Details about this subfamily can be found in Refs.~\cite{Creminelli:2017sry,Ezquiaga:2017ekz}. 
What matters for our analysis in the present paper is that this subfamily of DHOST theories generically
contains so-called beyond Horndeski terms.
In Horndeski theory, the non-linearities of the model normally screen the scalar field at small and intermediate scales, thus restoring the GR solution. This effect is a manifestation of the generic Vainshtein mechanism \cite{Vainshtein:1972sx,Babichev:2013usa}. The beyond Horndeski terms, on the other hand, lead to a breakdown of the Vainshtein mechanism inside matter. In particular, extra terms in the gravitational potentials appear \cite{Kobayashi:2014ida}. This feature has been used to get constraints on DHOST models from various physical observations, see e.g., \cite{Sakstein:2015aac,Jain:2015edg,Sakstein:2015zoa,Dima:2017pwp,Saltas:2018mxc}.
In this paper we point out that the presence of the beyond Horndeski terms may lead to pathological behaviours of the perturbations, 
generate instabilities in simple homogeneous backgrounds, 
and affect the every-day life physics, notably the speed of sound in fluids and gases.

The plan of the paper is as follows.
First, we derive in Sec.~\ref{sec:waves} an equation governing the propagation of waves in media, in the presence of beyond Horndeski terms. 
Then, we use these results to analyse and constrain the DHOST theories in Sec.~\ref{sec:constraints}. 
We conclude in Sec.~\ref{sec:conc}.

\section{Spherical sound waves in media and planar limit}
\label{sec:waves}


In  this section, we consider spherically symmetric ripples in some medium, in the framework of DHOST modified gravity.
As we will see, the modification of the gravitational potential with respect to GR affects standard hydrodynamics, 
and can lead to a pathological behaviour of the perturbations in the medium. 
The full equations in DHOST theories are rather complicated, due to the non-linear nature of these theories (The explicit form of the DHOST Lagrangian can be found in Appendix \ref{sec:DHOST}).
The main difficulty is that the theories remain non-linear even in situations when GR can be linearised. 
Non-linearity is an important feature of these theories, allowing to hide the fifth force generated by the extra degree of freedom $\varphi$. 
At the same time, this leads to technical difficulties in solving the full equations of motion.
In some cases, however, it is possible to solve approximately the system of equations. 

In our analysis, we will need the Newtonian potential for DHOST theories.
Let us focus in particular on the Newtonian potential $\Phi$ in the spherically symmetric case. Writing the metric as
\begin{equation}
ds^2=-[1+2\Phi(t,r)]dt^2+a(t)^2[1+2\Psi(t,r)](dr^2+r^2 d\Omega^2),
\end{equation}
$\Phi$ is determined by the following expression \cite{Kobayashi:2014ida,Koyama:2015oma,Saito:2015fza,Babichev:2016jom,Crisostomi:2017lbg,Langlois:2017dyl}:
\begin{equation}
\Phi'=\dfrac{G_\mathrm{N}(t)M(t,r)}{r^2}+\dfrac{\Upsilon_1(t)G_\mathrm{N}(t)}{4}M''(t,r),
\label{eq:PhibH}
\end{equation}
where a prime stands for a radial derivative, $G_\mathrm{N}(t)$ is the effective Newtonian constant, and $\Upsilon_1(t)$ a dimensionless parameter, the expression of which can be found in Appendix \ref{sec:DHOST}. $G_\mathrm{N}$ and $\Upsilon_1$ depend on the cosmological background, and hence on time. However, this time evolution is governed by the Hubble rate, and we can thus treat these parameters as constant over the time scales we are interested in. We will omit their explicit time dependence in what follows. 
$M(t,r)$ is the integral of the energy density $\rho$ over a ball of radius $r$:
\begin{equation}
M=4 \pi a(t)^2 \displaystyle\int r^2\,dr \rho(t,r).
\end{equation}
We 
do not give an expression for the other potential, $\Psi$, since we will be interested in Newtonian hydrodynamics, when $\Psi$ can be safely neglected. Similarly, the scale factor $a$ is determined through cosmology and can be set to unity for our purposes.

Equation (\ref{eq:PhibH}) will be very handy in what follows, because it determines the weight. Indeed, in the absence of external force, test-particles follow the geodesics of spacetime. In the Newtonian picture, this effect is accounted for by saying that particles experience weight. This force, per unit mass, is minus the 3-gradient of the Newtonian potential, $\vec{\nabla}\Phi$. In the case of spherical symmetry, the magnitude of the weight is thus proportional to $\Phi'$. 

In deriving Eq.~(\ref{eq:PhibH}), besides spherical symmetry, the following assumptions were made:
\begin{itemize} 
\item the gravitational potentials are small, $(\Phi,\Psi)\ll 1$, and the gradient of the scalar field is small with respect to its time derivative, $(\varphi'/\dot\varphi)^2\ll1$,
\item quasi-staticity,
\item Vainshtein regime, i.e., the canonical kinetic term ---~normally assumed to be present in the action~--- is subdominant compared to the non-linear terms.
\end{itemize}
The first condition is implied by a weak-field approximation, and should be satisfied in configurations where the curvature remains small, such as the Solar System.
The assumption on derivatives of $\varphi$ is technical and it can be checked a posteriori: it is needed to neglect terms $\mathcal{O}[(\varphi'/\dot\varphi)^4]$ with respect to $\mathcal{O}[(\varphi'/\dot\varphi)^2]$ when obtaining Eq.~(\ref{eq:PhibH}).
The second condition ---~quasi-staticity~--- is satisfied when the sound in matter propagates much slower than the spin-2 and spin-0 perturbations.
If one is interested in a theory with equal speeds of gravity and light, this condition is obviously correct for tensor modes.
For the scalar, the speed of waves depends on the particular choice of a DHOST model. A very specific self-tuning would however be required to make this speed as small as the speed of sound. We simply assume this is not the case, i.e., the scalar perturbations propagate with a much higher speed than matter waves. Under this assumption, the gravitational and scalar field can be determined through the static field equations.
Indeed, their evolution is driven by the slow evolution of matter density and pressure.
Both tensor and scalar modes ``see'' the matter waves as being almost static,
since the metric and scalar field almost immediately adjust to their equilibrium (and static) value, long before the matter density has time to evolve again. 

The last assumption can be, in fact, lifted, so that one does not assume that the Vainshtein regime is on. 
It can be checked, that without assuming the Vainshtein regime, one gets an extra term in Eq.~(\ref{eq:PhibH}), due to the canonical kinetic term $(\pd\varphi)^2$ in the Lagrangian\footnote{We, however, still assume the smallness of the gradient of the scalar field, although usually these two conditions are related.}:
\begin{equation}
\Phi'=\dfrac{G_\mathrm{N}M}{r^2}+\dfrac{\Upsilon_1G_\mathrm{N}}{4}M''+ \alpha_1 \dot\varphi^2 r^2,
\label{eq:PhibHcor}
\end{equation}
where $\alpha_1$ is a dimensionless coefficient which depends on the theory and a background solution.
We will see later that this last term contributes only to the static background solution, and plays no role in the propagation of waves. The first term in the right-hand side is the usual Newtonian source. 
It is identical to what is found in the Newtonian limit of GR (up to a redefinition of $G_\mathrm{N}$ in DHOST theories). 
The second term, on the other hand, is entirely specific to the modified gravity model under consideration. The aim of what follows is to show that this term can have dramatic consequences on elementary physics, like the propagation of sound waves in the air.
The third term also depends on parameters of the theory, and physically describes the back-reaction of the metric on the background scalar field. Let us now check how the modified Newtonian potential (\ref{eq:PhibHcor}) affects standard hydrodynamics. 
Neglecting friction, the local dynamics of a gas or a fluid is described by Euler equation:
\begin{equation}
\dfrac{\partial(\rho\vec{v})}{\partial t}+(\vec{v}\cdot\vec{\nabla})(\rho\vec{v})=-\rho\vec{\nabla}\Phi-\vec{\nabla} P,
\label{eq:Euler3D}
\end{equation}
where $P$ is the pressure of the gas, $\rho$ its density and $\vec{v}$ its speed. This equation is supplemented by the continuity equation:
\begin{equation}
\dfrac{\partial\rho}{\partial t}+\mathrm{div}(\rho \vec{v})=0.
\label{eq:contin3D}
\end{equation}
The latter equation asserts the conservation of the mass density. Equations (\ref{eq:Euler3D}) and (\ref{eq:contin3D}) can be obtained in the weak-field limit of the relativistic conservation equation $\nabla_\mu T^{\mu\nu}=0$, with $T^{\mu\nu}$ the energy-momentum tensor of the fluid. In order to close the system of equations, one needs a relation between $P$ and $\rho$. 
For small perturbations, it is enough to have such a relation at the linearised level, assuming that the wave constitutes a small perturbation of the rest solution. This rest solution is characterized by a pressure $P_0$, a density $\rho_0$ (possibly functions of $r$) and a vanishing velocity of the medium. We define the perturbation through:
\begin{align}
P&=P_0+\delta P,
\\
\rho&=\rho_0+\delta \rho.
\end{align}
For practical purposes, the compression and expansion of a gas when a wave propagates through can be regarded as isentropic. Thus, at the linearised level,
\begin{equation}
\dfrac{\delta P}{\delta\rho}\simeq \left.\dfrac{\partial P}{\partial\rho}\right|_S,
\label{eq:eos}
\end{equation}
$S$ being the entropy. The right-hand side quantity is known in the case of an ideal gas, for instance. At this stage, let us note that we only have the spherically symmetric version of $\vec{\nabla}\Phi$ at our disposal. Therefore, we will focus on a one dimensional and spherically symmetric problem. 
We linearise Eqs.~(\ref{eq:Euler3D}) and (\ref{eq:contin3D}) in this spherically symmetric framework. A derivative with respect to $t$ will be noted with a dot, and a prime stands for a derivative with respect to $r$. The first term in the right-hand side of Eq.~(\ref{eq:Euler3D}), $-\rho\vec{\nabla}\Phi$, then becomes
\begin{equation}
-\rho\Phi'=-\rho_0\left\{4\pi G_\mathrm{N}\left[\dfrac{1}{r^2}\displaystyle\int dr'\rho_0 r'^2+\dfrac{\Upsilon_1}{4}(r^2\rho_0)'\right] + \alpha_1 \dot\varphi^2 r^2\right\} -(\rho\Phi')|_\mathrm{lin}+\mathcal{O}(\delta\rho^2),
\end{equation}
where the term inside brackets is the background contribution, when no perturbation is present: this is just the usual weight (in the framework of the modified gravity theory we study). The first order correction (in terms of $\delta\rho$) due to the presence of the wave, $(\rho\Phi')|_\mathrm{lin}$, is given by
\begin{equation}
\begin{split}
(\rho\Phi')|_\mathrm{lin} &= 4\pi G_\mathrm{N}\left\{\delta\rho\left[\dfrac{1}{r^2}\displaystyle\int dr' r'^2\rho_0+\dfrac{\Upsilon_1}{4}(r^2\rho_0)'\right]\right.
\\
&\quad\left.+\rho_0\left[\dfrac{1}{r^2}\displaystyle\int dr' r'^2\delta\rho+\dfrac{\Upsilon_1}{4}(r^2\delta\rho)'\right]\right\},
\end{split}
\label{eq:Philin}
\end{equation}
Using the above notation, the linearized versions of Eqs.~(\ref{eq:Euler3D}) and (\ref{eq:contin3D}) can be written respectively as
\begin{align}
\rho_0 \dot{v}&=-(\rho\Phi')|_\mathrm{lin}-\delta P',
\label{eq:Eulerlin}\\
0&=\dot{\delta\rho}+\rho_0 v'.
\label{eq:continlin}
\end{align}
We can simplify Eq.~(\ref{eq:Philin}), assuming further that the wavelength of perturbations $\lambda$ is much smaller than the radius of the spherical wave $r$.
This always happens when, for example, a point-like (or spherical) source radiates spherical waves in a homogeneous medium. 
Then, any point which is far enough from the source is automatically in the regime $r\gg \lambda$.
One would expect that, in the limit $r/\lambda\to \infty$, the above equations yield the same result as if the calculation had been carried out in planar symmetry; however, this does not happen, as we will see below. 

In this short-wavelength (or large distance) limit, the dominant contribution in the right-hand side of Eq.~(\ref{eq:Philin}) comes from the last term, since it contains the derivative of the density perturbation, $\delta\rho'\simeq \lambda^{-1}\delta\rho$. We thus get
\begin{equation}
-(\rho\Phi')|_\mathrm{lin} \simeq -4\pi G_\mathrm{N}\frac{\Upsilon_1}{4}\rho_0 r^2 \delta\rho',
\label{eq:PhibHapprox}
\end{equation}
where the other terms were neglected, assuming 
\begin{equation}
\frac{r}\lambda\gg 1, \;\;\; \frac{r}\lambda\gg \frac{1}{\Upsilon_1}\frac{M}{\rho_0 r^3}.
\end{equation}
Note in passing that $-(\rho\Phi')|_\mathrm{lin}$, and thus $\Phi$, can be made as small as desired by choosing a small amplitude for $\delta\rho$. This way, we can always satisfy the first assumption in the above list. Combining Eqs.~(\ref{eq:eos}), (\ref{eq:Eulerlin}), (\ref{eq:continlin}) and (\ref{eq:PhibHapprox}), one can extract a wave equation on $\delta\rho$:
\begin{equation}
\ddot{\delta\rho}-\left(\left.\dfrac{\partial P}{\partial \rho}\right|_S+4\pi G_\mathrm{N}\dfrac{\Upsilon_1}{4}\rho_0 r^2\right)\delta\rho''=0.
\label{eq:deltarho}
\end{equation}
This means that, in the DHOST theory under consideration, sound waves propagate ---~locally, at radius $r$~--- at a speed $c_\mathrm{DHOST}$ such that
\begin{equation}
c_\mathrm{DHOST}^2=\left.\dfrac{\partial P}{\partial \rho}\right|_S+4\pi G_\mathrm{N}\dfrac{\Upsilon_1}{4}\rho_0 r^2.
\label{eq:csDHOST}
\end{equation}
This is to compare to the corresponding quantity in GR, where the Newtonian potential does not affect the speed of propagation:
\begin{equation}
c_\mathrm{GR}^2=\left.\dfrac{\partial P}{\partial \rho}\right|_S.
\end{equation}
Let us examine the expression (\ref{eq:csDHOST}) in more detail.
First of all, notice that in DHOST theories, the speed of sound strongly depends on the background (cosmological) solution through the field $\varphi$, which is encoded in $\Upsilon_1$.
Secondly, the speed of sound in media depends on the distance to the center of the mass distribution (or to the source of the perturbation, if one has in mind a homogeneous medium). Moreover, 
the absolute value of $c_\mathrm{DHOST}^2$ grows as $r^2$ at large distances. 
Finally, the planar limit $r/\lambda\to \infty$ of a spherical wave does not coincide with the result for a planar wave. 
Indeed, for a purely planar wave, the higher-order DHOST terms do not contribute to the equations of motion ---~they become trivial because of the high symmetry of the problem. Therefore, in this configuration, the speed of sound remains the same as in GR.
On the other hand, taking the planar limit of the spherical wave, $ r/\lambda\to \infty$, we can see that the speed of sound changes considerably, as follows from Eq.~(\ref{eq:csDHOST}). This sort of discontinuity may indicate a hidden pathology of the theory. This problem resembles the van Dam-Veltman-Zakharov (vDVZ) discontinuity in Pauli-Fierz massive gravity, leading to  different physical predictions from those of GR, even when the mass of the graviton is sent to zero. There is a difference, however: in our case, there is a discontinuity when different solutions are compared for the \textit{same} theory, while the vDVZ discontinuity arises when solutions of the Pauli-Fierz theory are compared to those of GR. Therefore, the problem we pointed out above for DHOST theories seems to be indeed connected to the theory itself, although we postpone deeper investigations for later.

\section{Constraints on the theory}
\label{sec:constraints}

On top of the theoretical difficulties evoked above, our results can be used straightforwardly to constrain DHOST theories.
We consider two particular cases: dust in a homogeneous universe and sound waves in the air. 

As a first example, let us focus on a homogeneous dust background with a time-dependent scalar field $\varphi$.
This situation is realised, in particular, in scenarios when the present-day acceleration is driven by beyond Horndeski (or DHOST) terms, and dark matter consists of weakly interacting particles.
For our purposes, we can ignore here the effects of cosmological expansion, so that the spacetime is effectively flat, which is a valid assumption for $r\ll H^{-1}$, where $H$ is the Hubble parameter.
By definition, for pressureless matter, $P=0$. Then it follows from Eq.~(\ref{eq:csDHOST}) that 
\begin{equation}
c_\mathrm{DHOST}^2=4\pi G_\mathrm{N}\dfrac{\Upsilon_1}{4}\rho_0 r^2.
\label{eq:csdust}
\end{equation}
This means, in particular, that negative values of $\Upsilon_1$ are excluded. Indeed, for negative $\Upsilon_1$, $c_\mathrm{DHOST}^2$ is negative as well, and a gradient instability appears at any $r$. Thus, a small perturbation would immediately grow and the dust would cluster, preventing the existence of homogeneous configurations. Note that, for a given wavelength, a larger $r$ implies a stronger gradient instability. 
This implies, in particular, that an instability would also arise for matter with non-zero pressure in the models with negative $\Upsilon_1$. In this case, one only needs to choose a sufficiently large $r$, so that the second term in Eq.~(\ref{eq:csDHOST}) dominates over the first one.

\begin{figure}[t]
\centering
\includegraphics[width=.8\textwidth]{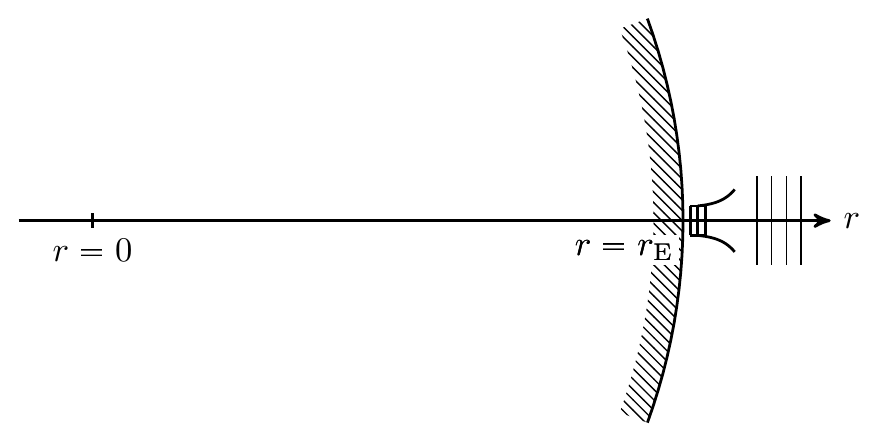}
\caption{Sound waves emitted radially outward by a loudspeaker at the surface of Earth, at a distance $r_\mathrm{E}$ of the origin.		
}
\label{fig:planspher}
\end{figure}

\begin{figure}[ht]
\centering
\includegraphics[width=\textwidth]{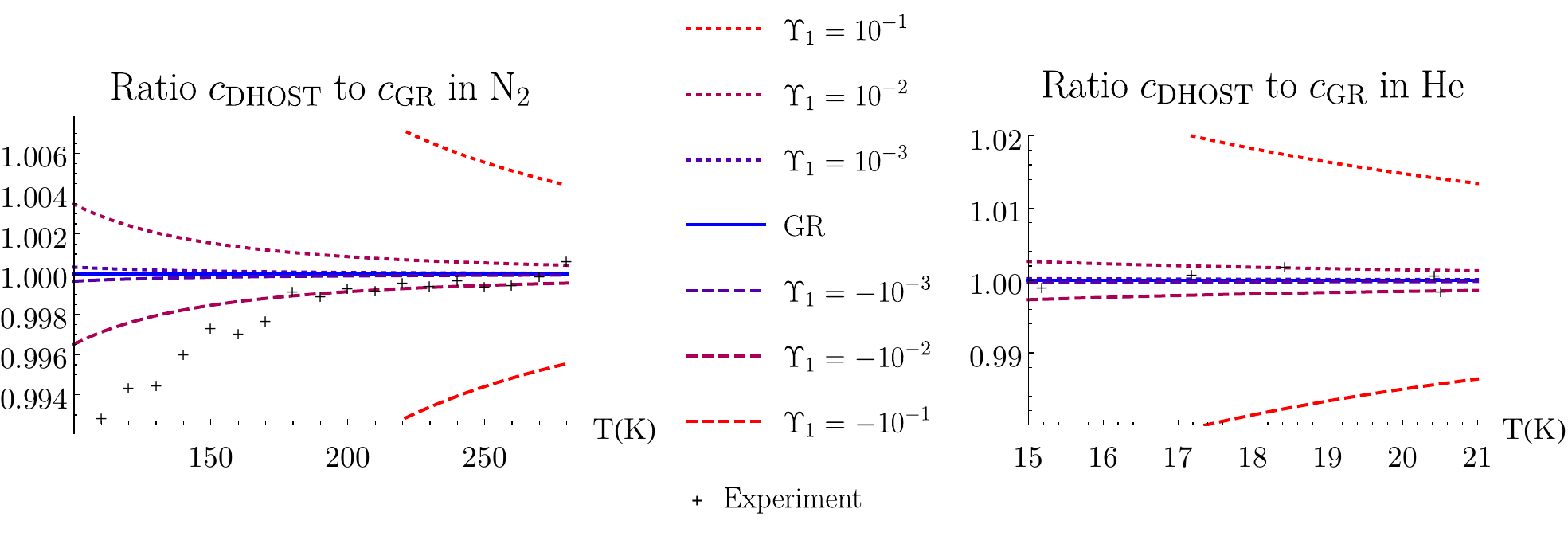}
\caption{Speed of sound as a function of the temperature in two different gases (nitrogen in the left panel, helium in the right one). All speeds are normalized to the value $c_\mathrm{GR}(T)$. The black crosses represent experimental highly accurate values \cite{jensen1980brookhaven}. The larger $|\Upsilon_1|$ is, the lesser the model agrees with the measurements. The low temperature deviation of measurements with respect to $c_\mathrm{GR}(T)$ in the left panel is due to the fact that, at these temperatures, nitrogen behaves less and less as an ideal diatomic gas.}
\label{fig:csvsT}
\end{figure}

As a second example, we consider the case of sound waves in the air, on Earth. 
It is interesting that a gravity modification of the DHOST type affects the everyday non-relativistic hydrodynamics.
Let us consider the situation of a vertically moving (almost) planar sound wave, as depicted in Fig.~\ref{fig:planspher}.
Since the radius of the Earth is much greater than the wavelength of the sound wave, we are in the regime described in the previous section. In particular, Eq.~(\ref{eq:csDHOST}) is valid in this case.
In the case of an ideal gas, the square of the speed of sound is $c_\mathrm{GR}^2=\gamma R T/M$, where $R$ is the ideal gas constant, $T$ is the temperature, $\gamma$ is the adiabatic index of the gas, and $M$ its molar mass. We take $r=r_\mathrm{E}\simeq 6.4\cdot10^6$~m for the radius of Earth, and atmospheric pressure $P\simeq10^5$~Pa. In the appropriate regime of temperature and pressure, the typical difference between the measured speed of sound and the prediction of the ideal gas model is  of order $0.2~\%$, or even less \cite{jensen1980brookhaven,zuckerwar2002handbook}. For a given DHOST theory, the deviation of $c_\mathrm{DHOST}$ with respect to $c_\mathrm{GR}$ is inversely proportional to the temperature $T$ (still in the case of an ideal gas). However, as temperature decreases, the model of an ideal gas is less and less accurate. In the range of validity of the ideal gas model, $c_\mathrm{DHOST}$ gives reasonable speed of sound values only as long as 
\begin{equation}
|\Upsilon_1|\lesssim10^{-2},
\label{eq:rangeU1}
\end {equation}
as illustrated in Fig.~\ref{fig:csvsT}.
For higher values of this parameter, there exists a significant deviation from the measurements. The range (\ref{eq:rangeU1}) is to be compared with ---~and improves~--- previously derived bounds \cite{Saito:2015fza,Sakstein:2015zoa,Dima:2017pwp,Saltas:2018mxc}. Such bounds were derived from astrophysical observations, the most accurate coming from white dwarves \cite{Saltas:2018mxc} (with the drawback that it is model dependent).  

Our intention here is not to present Eq.~(\ref{eq:rangeU1}) as a strong bound, because the speed of sound measurements of Refs.~\cite{jensen1980brookhaven,zuckerwar2002handbook} were not carried out exactly in the setup of Fig.~\ref{fig:planspher}. However, it is obvious in everyday-life physics that the speed of sound is locally isotropic. Thus, the precise setup of the experiments should not affect too much these estimates. We simply want to emphasize that very stringent constraints can be placed on DHOST theories, without involving astrophysics, through simple table-top experiments. The bound (\ref{eq:rangeU1}) could be highly improved by carrying out the calculation of the speed of sound out of spherical symmetry, and by taking into account more refined gas models, notably.

\section{Conclusion}
\label{sec:conc}
We explored how DHOST theories, considered as dark energy candidates, back-react on local physics. 
In our analysis this local effect is encoded in a single dimensionless parameter $\Upsilon_1$.
Because of the complexity of the equations in full DHOST theory, we were compelled to make a set of assumptions, all of which, however, are justified in realistic configurations, such as the Solar System.


We showed that the presence of a time-dependent cosmological scalar field in DHOST theories changes the propagation of waves in media, altering even present day non-relativistic hydrodynamics.
As a simple consequence, a spherical perturbation in a homogeneous medium filled with dust leads to an instability whenever $\Upsilon_1$ is negative. 
Thus, the case  $\Upsilon_1<0$ is excluded.
Moreover, the speed of sound, say in the atmosphere of the Earth, is significantly affected in a generic DHOST theory.
The model can reasonably account for measurements only when $|\Upsilon_1|\lesssim10^{-2}$. This bound can certainly be improved by more involved calculations, and a better modelling of the physical medium where sound propagates.

We should also mention that, when seen as effective field theories, DHOST theories can have a somewhat low energy cutoff scale. Reference \cite{deRham:2018red}, for instance, treated the case of Horndeski theory and estimates its cutoff frequency to be around 260~Hz, not much higher than the frequency of the GW170817 event. The constraint $\Upsilon_1>0$ is not affected by these considerations; one can always choose a large enough wavelength $\lambda$ for the perturbation, since spacetime is assumed to be homogeneous over a vast scale. The bound (\ref{eq:rangeU1}), however, is accurate only when the theory under consideration is not viewed as an effective field theory where higher-order corrections would enter at a frequency scale of a few hundred Hz.

More generically, the ill-behaved limit from spherical to planar symmetry seems to indicate a pathology of the theory. Together with the strong constraints imposed by the speed of gravitational waves, and by the non-decay of these waves \cite{Creminelli:2018xsv}, this is a strong hint against DHOST theories as dark energy candidates.

\acknowledgments{We thank Christos Charmousis and Marco Crisostomi for interesting discussions. 
E.B. acknowledges support from from the research program ``Programme national de cosmologie et galaxies'' of the CNRS/INSU, France, 
and from 
PRC CNRS/RFBR (2018--2020) n\textsuperscript{o}1985 ``Gravit\'e modifi\'ee et trous noirs: signatures exp\'erimentales et mod\`eles consistants''. 

\appendix

\section{DHOST theories}
\label{sec:DHOST}

The DHOST theories were fully investigated up to cubic order in the second derivatives of the scalar field \cite{Crisostomi:2016czh,Achour:2016rkg}. The most general DHOST Lagrangian can then be written
\begin{equation}
\label{dhost_action}
S_\mathrm{DHOST} =  \int d^4x \sqrt{-g} \left[ 
F_0(\varphi,X) + F_1(\varphi,X) \Box \varphi + F_2(\varphi,X) R + \sum_{I=1}^5 A_I(\varphi,X) L_I^{(2)} 
\right] \,,
\end{equation}
where $X=(\partial\varphi)^2$, $F_0$ and $F_1$ are free functions, while $F_2$ and the $A_I$ are free up to three degeneracy conditions that we give below; the Lagrangian densities $L_I$ read
\begin{align}
L_1^{(2)} &= \varphi^{\mu \nu}\varphi_{\mu \nu},
\\
L_2^{(2)} &=  (\Box\varphi)^2,
\\
L_3^{(2)} &=  \Box\varphi\, \varphi^\rho \varphi_{\rho \sigma} \varphi^{\sigma},
\\
L_4^{(2)} &=  \varphi^{\mu} \varphi_{\mu \nu}  \varphi^{\nu \rho } \varphi_{\rho}, 
\\
L_5^{(2)} &= ( \varphi^{\rho} \varphi_{\rho \sigma} \varphi^{\sigma})^2,
\end{align}
with $\varphi_{\mu \nu}=\nabla_\nu \nabla_\mu \varphi$, and $\varphi_\mu = \nabla_\mu \varphi$. Imposing the constraint that the speed of tensor modes and the speed of light are identical over cosmological backgrounds translates as $A_1=0$. The three degeneracy relations mentioned above then read
\begin{align}
A_2&=0,
\\
A_4&=-\dfrac{1}{8F_2}(8A_3F_2-48F_{2X}^2-8A_3F_{2X}X+A_3^2X^2),
\\
A_5&=\dfrac{A_3}{2F_2}(4F_{2X}+A_3X).
\end{align}
Once these relations are taken into account, the parameters introduced in Sec.~\ref{sec:waves} are given by:
\begin{align}
\dfrac{1}{8\pi G_\mathrm{N}(t)}&=2(F_2-XF_{2X})-\dfrac32A_3X^2,
\\
\Upsilon_1(t)&=-\dfrac{(4F_{2X}-XA_3)^2}{4A_3F_2}.
\end{align}
If $\varphi$ is chosen to have no mass dimension, and one introduces a unique mass scale $M$ in the scalar sector, this mass is expected to be very tiny, of the order of the Hubble rate $H$. The time derivative of $\varphi$ should then be of similar order. Thus, unless large numbers or unnatural cancellations are hidden in the functions $F_2$ and $A_3$, $\Upsilon_1$ is expected to be of order 1.

\bibliographystyle{unsrt}
\bibliography{biblio}

\end{document}